\documentclass[a4paper]{jpconf}
\usepackage{graphicx}

\begin{document}

\title{Evolution of fragmented states}

\author{G.~S.~Paraoanu}

\address{Low Temperature Laboratory, Helsinki University of Technology,
P.O. Box 5100, FIN-02015 TKK, Finland}

\ead{paraoanu@cc.hut.fi}

\begin{abstract}
We consider the problem of evolution of the many-body state of a weakly interacting
system of bosons in an initially fragmented (Fock) state. We show that the state at any time can be
expressed as a continuous superposition of an infinite number of Gross-Pitaevskii states.

PACS:03.75.-b, 03.65.-w, 03.75.Kk

\end{abstract}

Quantum mechanics incorporates in a single elegant formalism features that resemble wave-like
behavior and particle-like behavior. The many-body field operator $\hat\psi (x)$ for example, not only
describes the annihilation/creation of a particle at position $x$, but it comes as well with a phase,
as do classical fields describing wavelike phenomena. The phenomenology associates with the
quantum-mechanical phase is vast: narrowing down our focus to the literature related to superfluidity and superconductivity, we find that
phases play a crucial role in the description of the Josephson effect. Here we concentrate on the case
of atomic Bose-Einstein condensates, in which Josephson effects are by now theoretically and experimentally on solid ground
\cite{josephson}. In this case, the system consists of two condensates, and the relative phase between them is assumed
to be established previously, as in the case of metallic superconductors separated by an insulator, in which electrons can tunnel
between the two sides before and during the cooling process. With atomic Bose-Einstein condensates a new type
of experiment \cite{anderson} is possible, if two such gases are connected by a weak link or put to interfere only after they have been cooled separately
below the Bose-Einstein transition temperature. Now the initial state is a fragmented (Fock) state $\mid N/2,N/2\rangle_{t=0}$,
with no well-defined relative phase
between the two components.
In fact, one of the first experiments \cite{andrews} with condensed atomic gases has been along these directions, and a many others
followed \cite{new}. In the non-interacting case, the theory describing the measurement-induced phase coherence
is now well-established \cite{reviews}; also, information about correlations in low-dimensional quasicondensates can be obtained by this interference technique \cite{pol}.

But describing interfering interacting gases which are prepared independently remains a challenging problem.
We analyzed recently this situation in \cite{me}. In this paper, we give more details about the derivation of an approximate
simple expression for the many-body wavefunction for the evolving system. We will use the subscripts $L$ and $R$ to denote
condensates prepared initially in a left- and a right- well, separated by a high enough
potential barrier so that tunneling is not possible before an initial moment $t=0$, when the condensates are allowed to recombine.
The field operator is $\hat{\psi}(x)$, $x$ is the coordinate, and the initial state is a fragmented (Fock state),

\begin{equation}
\mid \frac{N}{2}, \frac{N}{2}\rangle_{t=0}=\frac{1}{(N/2)!}\hat{a}_{L}^{\dagger (N/2)}(t=0)
\hat{a}_{R}^{\dagger (N/2)}(t=0)\mid 0 \rangle
.\label{st}
\end{equation}
where $\hat{a}_{L,R} (t=0) = \int dx
 \Phi_{L,R}^{*}(x,t=0)\hat{\psi}(x)$.
At $t=0$, the trapping and the barrier potential is switched off; the Hamiltonian for $t>0$ becomes (with $g= 4\pi\hbar^{2}a/m,$ $a>0$)
\begin{equation}
\hat{H} = \frac{-\hbar^{2}}{2m}\int
dx\hat{\psi}^{\dagger}(x)\nabla^{2}\hat{\psi} (x) + \frac{g}{2}\int
dx\hat{\psi}^{\dagger}(x)\hat{\psi}^{\dagger}(x)\hat{\psi} (x)\hat{\psi}
(x). \label{hh}
\end{equation}
What is the structure of the many-body state at $t>0$? It is easy to convince oneself that this Hamiltonian does not preserve the
fragmented structure Eq. (\ref{st}) of the initial state. Another approach is then needed. We expand the fragmented state Eq. (\ref{st})
into phase states \cite{reviews}
\begin{equation}
\mid \frac{N}{2}, \frac{N}{2}
     \rangle_{t=0} =
     \frac{c_{0}}{\sqrt{2^{N}N!}}\int_{0}^{2\pi}\frac{d\varphi}{2\pi}\left[
     \hat{a}_{L}^{\dagger}(t=0)e^{i\varphi /2}+\hat{a}_{R}^{\dagger}(t=0)e^{-i\varphi
     /2}\right]^{N}\mid 0 \rangle ,
\end{equation}
where $c_{0} = 2^{N/2}(N/2)!/\sqrt{N!}$ \cite{reviews}.

With the notation
$
\mid \Phi_{\varphi }(t=0) \rangle_{N} =
(2^{N}N!)^{-1/2}\left[\hat{a}_{L}^{\dagger}(t=0)e^{i\varphi
/2}+\hat{a}_{R}^{\dagger}(t=0)e^{-i\varphi /2}\right]^{N}\mid 0\rangle
$
for the phase state, we have
\begin{equation}
\mid \frac{N}{2}, \frac{N}{2}
     \rangle_{t=0} =
     c_{0}\int_{0}^{2\pi}\frac{d\varphi}{2\pi}
      \mid \Phi_{\varphi }(t=0) \rangle_{N}.
\end{equation}
This form is not invariant either under the action of the evolution Hamiltonian.
However, we show that within the same degree of approximation as the standard time-dependent Gross-Pitaevskii
equation, the state at a later moment $t>0$ can be written as \begin{equation}
|\Psi \rangle_{t}\simeq c_{0}\int_{0}^{2\pi}\frac{d\varphi}{2\pi}
      \mid \Phi_{\varphi }(t) \rangle_{N}.
\end{equation}
So although a fragmented state does not have a single order parameter, as we will show below,
it can be approximately described by an infinity of order parameters
$\sqrt{N}\Phi_{\varphi }(x,t)$. This comes from the fact that the evolution Hamiltonian, in the limit
$N\rightarrow \infty$,
has approximately zero matrix elements between states $\mid\Phi_{\varphi }(t)\rangle_{N}$ with different
phases $\varphi$; therefore each of these states is evolved independently.
We start our proof by noticing that, as an immediate application of the %%@
Cauchy-Bunyakovski-Schwarz inequality in $L^{2}$ Banach spaces,
$\mid \langle \Phi_{\varphi '}(t)\mid \Phi_{\varphi} (t)\rangle \mid$ reaches a maximum %%@
value of 1 for $\varphi ' = \varphi$.
The quantity
$\mid _{N}\langle \Phi_{\varphi '}(t)\mid \Phi_{\varphi} (t)\rangle_{N} \mid$
will then be very close to zero for $\varphi ' \neq \varphi$, and strongly peaked
to 1 if   $\varphi ' = \varphi$, so we can write
\begin{equation}
\mid ~_{N}\langle \Phi_{\varphi '}(t)\mid \Phi_{\varphi} (t)\rangle_{N} \mid
\approx  1 - \frac{\pi N}{\alpha^{2}_{\varphi}(t)} (\varphi ' -\varphi )^{2} \approx \exp \left[-\frac{\pi N}{\alpha^{2}_{\varphi}(t)}(\varphi ' -\varphi)\right]
\approx \frac{\alpha(t)}{\sqrt{N}}
\delta (\varphi ' -\varphi),
\end{equation}
where   $ \alpha(t) = \sqrt{2\pi}\left\vert (d^{2}/d\varphi '^{2}) \mid \langle \Phi_{\varphi '}\mid \Phi_{\varphi} \rangle (t)
\mid_{\varphi ' = \varphi}\right\vert
^{-1/2}$.

We now introduce \cite{castin-gardiner}, at any time $t$ and for a given phase
$\varphi$, a projection operator $R_{\varphi }(t)$ orthogonal to the
subspace spanned by the ket $\mid\Phi_{\varphi }(t)\rangle$, $\hat{R}_{\varphi }(t)=
1-\mid\Phi_{\varphi}(t)\rangle\langle\Phi_{\varphi}
(t)\mid $. Then, at any time
$t$, we split the field operator into a condensate part of the order of $\sqrt{N}\times$wavefunction and smaller phonon part
of the order of 1$\times$wavefunction, $\hat{\psi} (x)=\hat{\Phi}_{\varphi}(x,t) + %%@
\hat{\chi}_{\varphi}
(x,t)$, where

\begin{equation} \hat{\chi}_{\varphi} (x,t) = \int dx'
R_{\varphi}(x,x';t)\hat{\Phi}_{\varphi}(x').
\end{equation}

It is straightforward to check that
$\hat{\chi}_{\varphi} (x,t)$ satisfies the commutation relations:
\begin{equation}
\left[\hat{\chi}_{\varphi} (x,t), \hat{\chi}_{\varphi} ^{\dagger}(x',t)\right] =
R_{\varphi}(x,x',t),
~~~~~~~~\left[\hat{\chi}_{\varphi} (x,t),
\hat{\Phi}_{\varphi}^{\dagger}(x',t)\right] = 0
.\label{c1} \end{equation}

In the Heisenberg picture, the evolution of the
phonon operator $\hat{\chi}^{(H)}(x,t)$
 (we put an extra superscript $(H)$ for operators written in the
Heisenberg representation) is given both by the Heisenberg commutator
and the intrinsic evolution due to the change of the modes on which
condensation occurs:
\begin{equation} i\hbar \frac{\partial}{\partial
t}\hat{\chi}_{\varphi}^{(H)}(x,t) = \left[\hat{\chi}_{\varphi}^{(H)}(x,t), \hat{H}\right] %%@
+
i\hbar
\left[\frac{\partial}{\partial t}\hat{\chi}_{\varphi}
(x,t)\right]^{(H)}. \label{1}
\end{equation}
We proceed now to the next step, which is
to linearize the Hamiltonian (\ref{hh}) around  $\hat{\Phi}_{\varphi}$
$\hat{\Phi}_{\varphi}^{\dagger}$,
\begin{eqnarray}
\hat{H} &=& -\frac{\hbar^{2}}{2m}\int dx
\hat{\Phi}_{\varphi}^{\dagger}(x,t)\nabla^{2}\hat{\Phi}_{\varphi}(x,t) +
\frac{g}{2}\int dx
\hat{\Phi}_{\varphi}^{\dagger}(x,t)\hat{\Phi}_{\varphi}^{\dagger}(x,t)\hat{\Phi}
_{\varphi}(x,t)\hat{\Phi}_{\varphi}(x,t)
\nonumber\\
 & & -\frac{\hbar^{2}}{2m}\int dx
\hat{\chi}^{\dagger}_{\varphi}(x,t)\nabla^{2}\hat{\Phi}_{\varphi}(x,t) + g\int dx
\hat{\chi}
^{\dagger}_{\varphi}(x,t)\hat{\Phi}_{\varphi}^{\dagger}(x,t)\hat{\Phi}_{\varphi}(x,t)\hat{\Phi}_{\varphi}(x,t)\nonumber
\\ &
& -\frac{\hbar^{2}}{2m}\int dx
\hat{\Phi}_{\varphi}^{\dagger}(x,t)\nabla^{2}\hat{\chi}_{\varphi}(x,t) + g\int dx
\hat{\Phi}_{\varphi}^{\dagger}(x,t)\hat{\phi}_{\varphi}^{\dagger}(x,t)\hat{\Phi}
_{\varphi}(x,t)\hat{\chi}_{\varphi}
(x,t) \nonumber
\\ &  &
+ .... {\cal O}(\hat{\chi}_{\varphi}\hat{\chi}_{\varphi} ,
\hat{\chi}^{\dagger}_{\varphi}\hat{\chi}^{\dagger}_{\varphi},
\hat{\chi}^{\dagger}_{\varphi}\hat{\chi}_{\varphi} ), \nonumber\end{eqnarray}
then calculate the commutator, and, keeping
only
the higher-order contribution (in $\sqrt{N}$), we get
\begin{equation}
i\hbar\frac{\partial}{\partial t} \hat{\chi}_{\varphi} (x,t) =
i\hbar\int dx \dot{R}_{\varphi}(x,x';t)\hat{\psi} (x') \simeq i\hbar\int dx'
\dot{R}_{\varphi}(x,x',t)\hat{\Phi}_{\varphi}(x',t).
\end{equation}
But $\int
R_{\varphi}(x,x',t)\hat{\Phi}_{\varphi}(x',t) = 0$, therefore taking a time-derivative we obtain
for the intrinsic time-dependence the formula
\begin{equation}
i\hbar\left[\frac{\partial}{\partial t}\hat{\chi}_{\varphi}
(x,t)\right]^{(H)}=\int
dx' R_{\varphi}(x,x',t)\left[-i\hbar\frac{\partial}{\partial
t}\hat{\Phi}_{\varphi}(x',t)\right]^{(H)}.
\end{equation}
For the usual
Heisenberg commutator of (\ref{1}) we use the commutation relations
(\ref{c1}) and the expansion form of the Hamiltonian and obtain
\begin{equation}
\left[\hat{\chi}_{\varphi}^{(H)}(x,t), \hat{H}\right] = \int dx'
R_{\varphi}(x,x';t)\left[ \frac{-\hbar^{2}}{2m}\nabla^{2}\hat{\Phi}_{\varphi}(x',t) +
g\hat{\Phi}_{\varphi}^{\dagger}(x',t)\hat{\Phi}_{\varphi}(x',t)\hat{\Phi}_{\varphi}(x',t)
\right]^{(H)} .\end{equation}
As a result, we finally get for (\ref{1}):
\begin{eqnarray} i\hbar\frac{\partial}{\partial t}\hat{\chi}_{\varphi}^{(H)}(x,t) =
\int
dx'R_{\varphi}(x,x';t)\left[ -i\hbar\frac{\partial}{\partial
t}\hat{\Phi}_{\varphi}(x',t) -\frac{\hbar^{2}}{2m}\nabla^{2}\hat{\Phi}_{\varphi}(x',t) +
g\hat{\Phi}^{\dagger}_{\varphi}(x',t)\hat{\Phi}_{\varphi}(x',t)\hat{\Phi}
_{\varphi}(x',t)\right]^{(H)}.
\nonumber
\end{eqnarray}

The consistency condition for the expansion $\hat{\psi} (x) =
\hat{\Phi}_{\varphi}(x,t)
+ \hat{\chi}_{\varphi} (x,t)$ is \begin{equation}
\int d\varphi ' _{N-1}\langle \Phi_{\varphi '}(t)\mid \hat{\Phi}_{\varphi}(x,t)\mid\Phi_{\varphi }(t)\rangle_{N}  \gg
\int d\varphi '_{N-1}\langle\Phi_{\varphi '}(t)\mid\hat{\chi}_{\varphi} (x,t)\mid %%@
\Phi_{\varphi '}(t)\rangle_{N}, \label{olo}\end{equation}
or in other words the quantum depletion of a state $\varphi$ should not acquire %%@
contributions of the order of the corresponding order parameter during evolution:
\begin{eqnarray}
&i\hbar \frac{d}{dt}\int d\varphi ' _{N-1}\langle \Phi_{\varphi '}(t)\mid\hat{\chi}_{\varphi}(x,t) \mid \Phi_{\varphi }(t)\rangle_{N} \approx \int d\varphi ' _{N-1}\langle \Phi_{\varphi '}(0)\mid\int
dx'R_{\varphi}(x,x';t) \left[-i\hbar\frac{\partial}{\partial
t}\hat{\Phi}_{\varphi}(x',t) \right.& \nonumber \\
 &\left.-\frac{\hbar^{2}}{2m}\nabla^{2}\hat{\Phi}_{\varphi}(x',t)+
g\hat{\Phi}_{\varphi}^{\dagger}(x',t)\hat{\Phi}_{\varphi}(x',t)\hat{\Phi}_{\varphi}(x',t)\right]
^{(H)}\mid \Phi_{\varphi}(0)\rangle_{N}
= \int dx'R_{\varphi}(x,x';t)\int d\varphi ' _{N-1}\langle \Phi_{\varphi '}(t)\mid
& \nonumber \\
&\left[-i\hbar\frac{\partial}{\partial t}\hat{\Phi}_{\varphi}(x',t)
-\frac{\hbar^{2}}{2m}\nabla^{2}\hat{\Phi}_{\varphi}(x',t) +
g\hat{\Phi}_{\varphi}^{\dagger}(x',t)\hat{\Phi}_{\varphi}(x',t)\hat{\Phi}_{\varphi}(x',t)\right]
\mid \Phi_{\varphi}(t)\rangle_{N} &\nonumber \\
&\simeq \alpha_{\varphi}(t)\int dx'R_{\varphi}(x,x';t)\left[-i\hbar\frac{\partial}{\partial t}\Phi_{\varphi}(x',t) -\frac{\hbar^{2}}{2m}\nabla^{2}\Phi_{\varphi}(x',t) +
gN\mid\Phi_{\varphi}(x',t)\mid^{2}\Phi_{\varphi}(x',t)\right]. &\nonumber
\end{eqnarray}
Thus, to obtain the approximate
evolution for the fragmented states which satisfies Eq. (\ref{olo}) we
have to impose a time-dependent Gross-Pitaevskii evolution for each wavefunction $\Phi_{\varphi}(x,t)$,
\begin{equation}
i\hbar\frac{\partial}{\partial t}\Phi_{\varphi}(x',t) =
-\frac{\hbar^{2}}{2m}\nabla^{2}\Phi_{\varphi}(x,t) +
gN\mid\Phi_{\varphi}(x,t)\mid^{2}\Phi_{\varphi}(x,t) .\label{gp}
\end{equation}

It is also easy to convince oneself, that, by the same procedure, one can show that
the approximation works for any many-body operator constructed from $\hat{\psi}$. Indeed, in the second quantization
any number-conserving
operator can be written as  $\hat{{\cal O}}=\sum_{k}{\cal O}_{k}[\hat{\psi}^{\dagger}, \hat{\psi}]$, where
${\cal O}_{k}[\hat{\psi}^{\dagger}, \hat{\psi}]$ is a functional in which each operator $\hat{\psi}^{\dagger}$ and $\hat{\psi}$ enters $k$ times.
Then we use $\hat{\psi} (x)=\hat{\Phi}_{\varphi}(x,t) +
\hat{\chi}_{\varphi}(x,t)$, we expand in powers of   $\hat{\chi}_{\varphi}(x,t)$
and, by following a similar procedure as above, we calculate $i\hbar (d/dt) _t\langle \Psi|
{\cal O}_{k}[\hat{\psi}^{\dagger}, \hat{\psi}] |\Psi\rangle_t$ and show that the first-order contribution in   $\hat{\chi}_{\varphi}(x,t)$
vanishes, if  Eq. (\ref{gp}) is satisfied. Therefore, in the thermodynamical limit and given that $k\ll N$, we obtain
\begin{equation}
_t\langle \Psi|  \hat{\cal O}|\Psi\rangle_t \simeq
  \frac{c_{0}^2}{4\pi^2}\sum_{k}N^{\frac{2k-1}{2}}\int d\varphi
  \alpha_{\varphi}(t){\cal O}_k[\Phi_{\varphi}^{*}(t),\Phi_{\varphi}(t)].
   \end{equation}

For example, if we consider the kinetic energy operator $-(\hbar^2 /2m)\int dx \hat{\psi}^{\dagger}(x)\nabla^{2}
\hat{\psi}(x)$, we find in the same way, by expanding the field operator, that
\begin{equation}
_t\langle \Psi| -\frac{\hbar^2}{2m}\int dx \hat{\psi}^{\dagger}(x)\nabla^{2} \hat{\psi}(x) |\Psi\rangle_t  \simeq
-\frac{c_{0}^2\sqrt{N}\hbar^2}{8\pi^2 m}\int \int dx d\varphi
   \alpha_{\varphi}(t)\Phi_{\varphi}^{*}(x,t)\nabla^{2}\Phi_{\varphi}(x,t).
\end{equation}

In conclusion, we show that the evolution of a fragmented (Fock) state of two weakly interacting Bose-Einstein condensed atomic gases
can be approximated by a many-body state consisting of an infinite superposition of Gross-Pitaevskii states with diferent initial conditions,
corresponding to different initial phase differences between the condensates.
\ack

This work was supported by the Academy of Finland (Acad. Res. Fellowship 00857 and %%@
projects 7111994 and 7118122).

\end{document}